\documentclass[useAMS,usenatbib]{mn2e}
\usepackage{epsfig}
\newcommand{\mnras}{MNRAS}
\newcommand{\apj}{APJ}
\newcommand{\apjsub}{APJ submitted}
\newcommand{\apjs}{APJS}
\newcommand{\apjl}{APJL}
\newcommand{\aap}{A\&A}
\newcommand{\aj}{AJ}
\newcommand{\nat}{Nature}
\newcommand{\araa}{ARA\&A}
\title[Dry Mergers]
{Dry Mergers: A Crucial Test for Galaxy Formation}
\author[Khochfar \& Silk]
  {S.~Khochfar,$^{1,2}$\thanks{sadeghk@mpe.mpg.de}
  J.~Silk,$^{1}$ \\
  $^1$ Department of Physics, Denys Wilkinson Building, Keble Road, Oxford OX1 3RH, United 
Kingdom\\
$^2$ Max Planck Institut f\"ur extraterrestrische Physik, p.o. box
1312, D-85478 Garching, Germany \\
   }
\date{Released 2008 Xxxxx XX}

\pagerange{\pageref{firstpage}--\pageref{lastpage}} \pubyear{2008}

\def\LaTeX{L\kern-.36em\raise.3ex\hbox{a}\kern-.15em
    T\kern-.1667em\lower.7ex\hbox{E}\kern-.125emX}

\begin{document}

\label{firstpage}

\maketitle
\begin{abstract}
We investigate the role that dry mergers play in the build-up of massive galaxies within the cold dark matter paradigm. Implementing an empirical shut-off mass scale for star formation, we find a nearly constant dry merger rate of $ \sim 6 \times 10^{-5}$ Mpc$^{-3}$ Gyr$^{-1}$ at $z \leq 1$ and a steep decline at larger z. Less than half of these mergers are between two galaxies that are morphologically classified as early-types, and the other half is mostly between an early-type and late-type galaxy. Latter are prime candidates for the origin of tidal features around red elliptical galaxies. The introduction of a transition mass scale for star formation has a strong impact on the evolution of galaxies, allowing them to grow above a characteristic mass scale of $M_{*,c} \sim 6.3 \times 10^{10}$ M$_{\odot}$ by mergers only. As a consequence of this transition, we find that around $M_{*,c}$, the fraction of 1:1 mergers is enhanced with respect to unequal mass major mergers. This suggest that it is possible to detect the existence of a transition mass scale by measuring the relative contribution of equal mass mergers to unequal mass mergers as a function of galaxy mass. The evolution of the high-mass end of the luminosity function is mainly driven by dry mergers at low z. We however find that only $10\% -20\%$ of galaxies more massive than $M_{*,c}$ experience dry major mergers within their last Gyr at any given redshift $z \le 1$.

\end{abstract}


\section{Introduction}
Mergers are fundamental to the cold dark matter paradigm of structure formation. They not only drive mass evolution by merging smaller dark matter haloes into larger ones, but they also change the morphology of galaxies 
from late to early-type \citep[e.g.][]{1972ApJ...178..623T,1992ARA&A..30..705B,2003ApJ...597..893N}, and drive gas to the centre of the merger remnant that triggers star formation \citep{1991ApJ...370L..65B,2006MNRAS.372..839N} and AGN activity \citep{2005Natur.433..604D}. Early structural studies of elliptical galaxies by \citet{1992ApJ...399..462B} already hinted at the mass-dependent importance of dissipation during their formation. In a first systematic study of the morphology of merging pairs in a CDM galaxy formation model, \citet{2003ApJ...597L.117K} could show that massive elliptical galaxies are mainly formed from dry mergers of early-type galaxies, while less massive ones show mixed mergers between an elliptical and a spiral galaxy. Only elliptical galaxies well below $L_*$ are predominantly formed by wet mergers from two spiral galaxies. Subsequent work on the role of dry mergers revealed that they can explain the formation of slow rotating boxy ellipticals 
\citep{2005MNRAS.359.1379K,2006ApJ...636L..81N}, that they lie on the fundamental plane \citep{2001ApJ...552L..13C,2003MNRAS.342..501N,2006MNRAS.369.1081B,2006ApJ...641...21R}, follow the $M_{\bullet}-\sigma$-relation \citep{2008arXiv0802.0210J} and that they could possibly explain the formation of a stellar density core in the centre of the remnant due to a binary black hole merger \citep{2001ApJ...563...34M,2004ApJ...613L..33G,2006ApJ...648..976M}. Furthermore, it has been argued that the strong size evolution of  massive early-type galaxies \citep{2007MNRAS.382..109T,2007ApJ...671..285T,2008ApJ...677L...5V,2008A&A...482...21C} provides evidence for dry merging \citep{2006ApJ...648L..21K}. In an attempt to model the size-evolution of early-type galaxies, \citet{2006ApJ...648L..21K} showed that the amount of dissipation during mergers can account for the observed size evolution. In their model, dry mergers result in remnants with larger sizes than remnants from gaseous mergers of the same mass.  Similar results have been reported from numerical simulations of mergers with varying degrees of gas fractions by \citet{2006ApJ...650..791C}. 

The natural question that immediately arises is, what is the reason for dry merging? The early seminal work of \citet{1977ApJ...215..483B}, \citet{1977ApJ...211..638S} and \citet{1977MNRAS.179..541R} predicts the existence of a characteristic mass scale, below which the cooling time $t_{cool}$ of a collapsing gas cloud is shorter than its dynamical  time $t_{dyn}$, allowing for efficient collapse on a dynamical time scale and subsequent star formation. In massive dark matter halos  with $M_{DM} > 10^{12}$ M$_{\odot}$, one generally finds $t_{cool} \gg t_{dyn}$ and that shock heating of the collapsing gas supports the formation of  a hot, static atmosphere at the halo virial temperature \citep[e.g.][]{2003MNRAS.345..349B,2007MNRAS.380..339B}. From an observational point of view, the existence of a bimodality in the properties of the galaxy population, occurring at a  mass scale of M$_* > 3 \times 10^{10}$ M$_{\odot}$ \citep{2003MNRAS.341...54K}, lends support to the notion of a transition in the mode of galaxy formation \citep{2006MNRAS.368....2D}. Hence one expects that dry merging will occur when cooling is sufficiently hindered at masses above a transition mass scale and if the reservoir of cold gas in the galaxy is used up by star formation before the merger happens. 

The existence of a characteristic shut-off mass scale in galaxy formation seemingly provides a simple way to truncate star formation
within galaxy formation models. Such an ad-hoc prescription has been used in earlier work by \citet{1999MNRAS.303..188K} with the aim of avoiding too massive and too blue galaxies in clusters, and more recently in work by \citet{2006MNRAS.370.1651C}. These latter  authors assume a shut-down of star formation in halos of mass $\geq 10^{12}$ M$_{\odot}$ at $z \leq 3$, and show  that the colour bimodality and luminosity function can be reproduced accurately in their model. The choice of $10^{12}$  M$_{\odot}$ draws its support from two main arguments laid out in \citet{2006MNRAS.368....2D}. 
One is that at this mass scale,  stable shocks appear that allow for shock heating of gas \citep{2003MNRAS.345..349B}. The second argument, more important, is that this shock-heated gas is generally so dilute and vulnerable to feedback that it literally stays hot forever and does not cool down to subsequently fuel star formation. While the first argument draws support from various simulations \citep[e.g.][]{2005MNRAS.363....2K}, the second argument is less clear. One main uncertainty is with regards to the heating source of the hot gas. Several plausible candidates are suggested in the literature such as  AGN-feedback \citep{1998A&A...331L...1S}, dynamical friction heating \citep{2004MNRAS.354..169E,2007ApJ...658..710N}, or heating by gravitational potential energy \citep{2008ApJ...680...54K,2008MNRAS.383..119D}. Neither the relative contribution nor the overall magnitudes with which these processes heat the hot gas are theoretically certain or observationally confirmed to date. 

The aim of this letter is twofold. Firstly, we predict the dry merger rate and its evolution by adopting a shut-off mass scale, and secondly, we use these results to propose observational strategies on how to test the existence  of a critical mass scale for the quenching of star formation, that relies on the continued merging activity within CDM-cosmologies and is independent to first order on the underlying baryonic physics involved in the quenching process.
\begin{figure}
\includegraphics[width=0.45\textwidth]{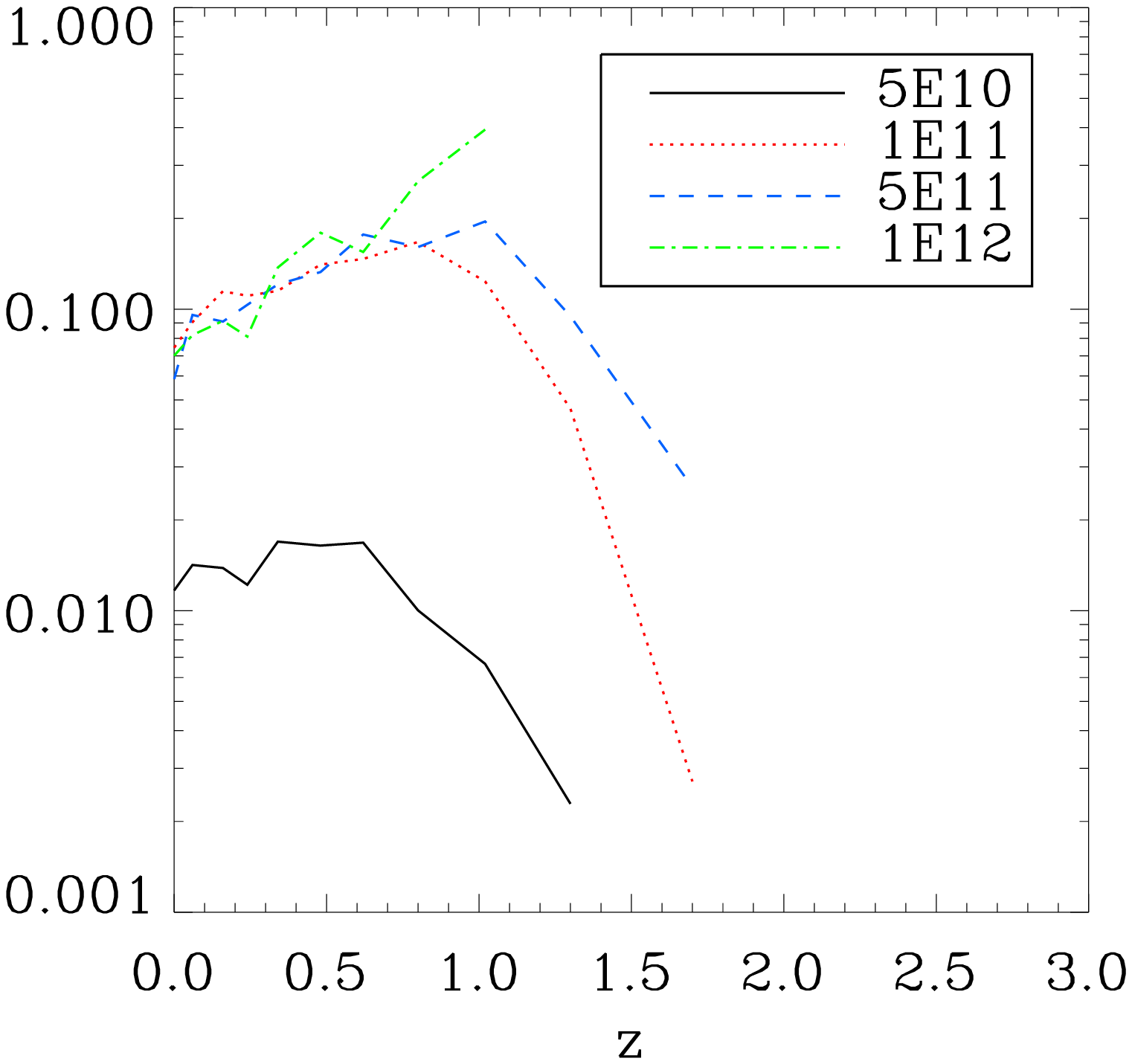}
 \caption{The fraction of galaxies that were formed by a dry major merger within the last Gyr. Lines show results for galaxy mass ranges $5 \times 10^{10} -10^{11}$, $10^{11} -5 \times 10^{11}$ M$_{\odot}$, $5 \times 10^{11} -10^{12}$ M$_{\odot}$ and $ >  10^{12}$ M$_{\odot}$. Galaxies with $M_* > M_{*,c}=6.3 \times 10^{10}$ M$_{\odot}$ show similar fractions of dry mergers at $z \le 1$. } \label{fig1}
\end{figure}

\section{The Model}\label{mod}
We use semi-analytical modeling (SAM) of galaxy formation to investigate the effect of the shut-off mass scale for cooling on the galaxy population. The dark matter history is calculated using the merger tree proposed by \citet{som99} with a mass resolution of $2 \times 10^9 M_{\odot}$. The baryonic 
physics within these dark matter haloes is calculated following recipes 
presented in \citet[][and references therein]{spr01}, including a model for the reionizing background 
by \citet{som02}. In our simulation, we assume that elliptical galaxies 
form whenever a major merger ($M_1 /M_2 \leq 3.5$ with $M_1 \geq M_2$) takes 
place. We assume that during this process, all the cold gas in the
 progenitor discs will be consumed in  a central starburst, adding to the 
spheroid mass, and that all stars in the progenitor discs will  
contribute to the spheroid as well. Furthermore, we also add the stars of satellite galaxies involved in  minor mergers to the spheroid. The merger time scale for galaxies is calculated using the dynamical friction prescription in \citet{spr01} and we find that the predicted merger rate is in good agreement with observations \citep{2001ApJ...561..517K,jogee}.
For more modeling details, we refer the reader to \citet{2005MNRAS.359.1379K} and \citet[KS]{ks06}. Throughout this paper, we use the following set of cosmological parameters derived from a combination of the 5-year WMAP data with Type Ia supernovae and measurements of baryon acoustic oscillations \citep{2008arXiv0803.0547K}:
$\Omega_0=0.28$, $\Omega_{\Lambda}=0.72$, $\Omega_b/\Omega_0=0.16$, 
$\sigma_8=0.8$ and $h=0.7$. 

In the following we will modify our fiducial model as laid out in SK by adopting a quenching of cooling in dark matter haloes above a critical mass scale of  $M_{DM,crit}\geq 10^{12}$ M$_{\odot}$ at $z \leq 3$ as suggested in \citet{2006MNRAS.368....2D}. Note that we allow the gas that is already in the disc to continue forming stars until it is used up, even after the host halo crossed $M_{DM,crit}$.  In the following, we define as dry mergers, 
objects for which M$_{gas,tot}/($M$_{*,tot}+$M$_{gas,tot})< 0.1$, with M$_{gas,tot}$ as the total amount of cold gas in both progenitor discs, and  M$_{*,tot}$ as the total amount of stars in both progenitors, respectively. Whenever we refer to dry mergers in the following we will mean, dry major mergers with $M_1 /M_2 \leq 3.5$ and $M_1 \geq M_2$. 
\begin{figure}
\includegraphics[width=0.45\textwidth]{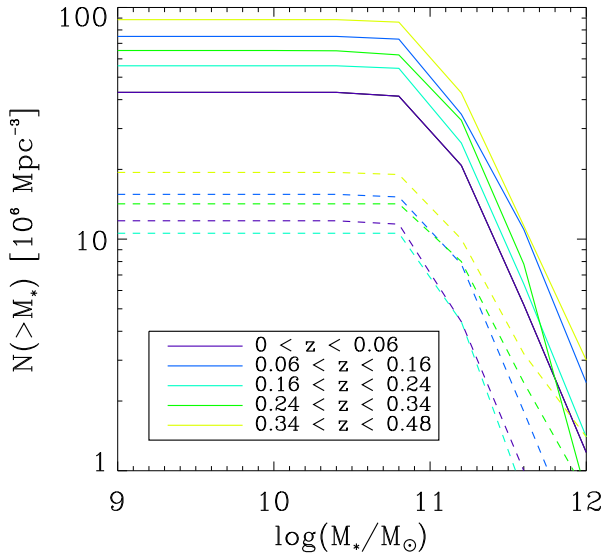}
 \caption{The cumulative number densities of dry mergers as a function of redshift and remnant stellar mass. Solid lines are prediction for the model with shut-off mass scale and dashed lines for the model without.}\label{fig4}
\end{figure}
 
\section{Evolution of Massive Galaxies} 
One of the main features of a shut-off mass scale $M_{DM,crit}$ is its influence on the evolution of massive galaxies that live in halos above $M_{DM,crit}$.
In a first implementation \citet{2006MNRAS.370.1651C} could show that they were able to reproduce the high mass tail of the luminosity function \citep{2003ApJ...592..819B} and hence prevent the common problem of overproducing too massive galaxies within SAMs. While the low-mass tail of the luminosity function becomes steeper with redshift  \citep{2007ApJ...668L.115K} independent of a shut-off, the high-mass tail of the luminosity function shows a much weaker evolution with time in models with shut-off, due to merging being the sole mode of growth compared to merging and star formation in our fiducial model without shut-off. In Fig. \ref{fig1} we show the fraction of massive galaxies that grew by dry major mergers within the last Gyr. In general the contribution from dry mergers decreases at $z>1$ as galaxies of the same mass tend to live in smaller dark matter haloes that fall below $M_{DM,crit}$. At redshifts $z<1$,  we find that in the mass range where dry mergers are significant, i.e. $ M_* > M_{*,c} \sim 6.3 \times 10^{10}$ M$_{\odot}$, the fraction of galaxies that where formed by a dry merger within the last Gyr is between $10 \%$ at $z=0$ and $20 \%$ at $z=1$ independent of the galaxy mass.  

\section{The Dry Merger Rate}
One critical point is the frequency of dry mergers in the universe. Observationally, there is still a vigorous debate going on as to whether dry mergers do not play any role \citep{2007ApJS..172..494S}, a mild role \citep{2007ApJ...654..858B}, or an important role \citep{2007ApJ...665..265F} in the growth of the most massive galaxies. The strategies to determine the influence of dry merging remain mostly centered on the evolution of the luminosity function and the colour bimodality of galaxies. In Fig. \ref{fig4}, we show the cumulative co-moving number density of dry major mergers as a function of galaxy mass in units of Mpc$^{-3}$. We calculated this number density by counting all dry major mergers that occurred within the cited redshift intervals.  The contribution to dry mergers comes mainly from galaxies around $M_{*,c}$. Galaxies more massive  than $M_{*,c}$ do not contribute significantly. The number densities increase by a factor of $\sim 2.5$ from $z=0$ to $z=0.34$ for galaxies more massive than $M_{*,c}$. We also show results for a model without shut-off mass scale in the same figure. The dry merger rates we find in this model are almost a factor 5 lower compared to the shut-off model. 
\begin{figure}
\includegraphics[width=0.45\textwidth]{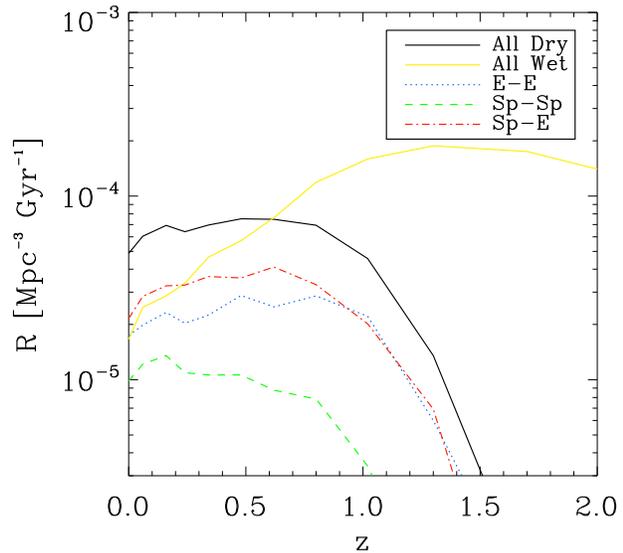}
 \caption{The merger rate as a function of redshift. Solid black and yellow lines show the overall dry and wet merger rates, respectively. The dashed and dotted lines divide the dry merger sample into sub-samples based on the bulge-to-total mass of the merging galaxies.  
}\label{fig2}
\end{figure}

\begin{figure}
\includegraphics[width=0.45\textwidth]{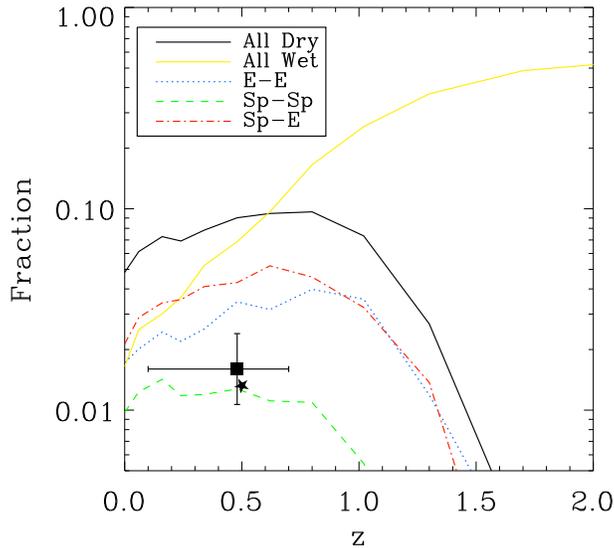}
 \caption{The merger fraction of galaxies for the same selections as in Fig. \ref{fig2}. The filled star and square are the modeled and observed dry merger fraction, respectively, for a sample of early-type galaxies as defined in \citet{2006ApJ...640..241B}.  }\label{fig3}
\end{figure}

To further quantify the evolution of massive galaxies in terms of dry mergers, we show the corresponding merger rates and fractions for galaxies more massive than $M_{*,c}$ in Fig. \ref{fig2} \& \ref{fig3}, respectively. We measure the dry major merger rate in our model by counting all dry mergers that occurred in our simulation volume within the last 1.0 Gyr of galaxies that are more massive than $M_{*,c}$. This gives the merger rate $R$ in units of Gyr$^{-1}$ Mpc$^{-3}$. The merger fraction $f$ is then calculated by simply dividing $R$ by the number density of galaxies more massive than $M_{*,c}$ at $z$.
 To make a consistent comparison to earlier observational work of \citet{2006ApJ...640..241B}, we calculate the fraction of dry mergers at $z=0.5$ by counting all dry mergers that galaxies with $M_*  \ge 10^{10} $M$_{\odot}$ in the last 150 Myr experienced  and dividing it by the number density of elliptical galaxies with $M_*  \ge 10^{10} $M$_{\odot}$. We define here and in the following elliptical galaxies as galaxies with bulge-to-total mass ratios of $B/T > 0.6$. For our comparison we only consider dry mergers between early-type galaxies and mixed mergers. It is likely that latter are contaminating the observed sample of \citet{2006ApJ...640..241B} and we find that in general more than half of all dry mergers are morphologically mixed mergers. As can be seen from Fig. \ref{fig3} the model output agrees well with the observations. 
Furthermore, we find an almost constant  dry merger rate at $z \le 1$ with $ \sim 6 \times 10^{-5}$ Gyr$^{-1}$ Mpc$^{-3}$ which shows a weaker decline to lower redshifts than the wet  merger rate.
The dry merger rate in general declines strongly at $z>1$ and is two orders of magnitude smaller than wet mergers at $ z \sim 1.5$. We continue by splitting up the sample of dry mergers based on the morphologies of the merging galaxies. Here we define galaxies with bulge-to-total stellar mass greater than 0.6 as ellipticals and all other galaxies as spirals. The relative contribution of different types of dry mergers to the merger fraction and rate is roughly constant throughout time. The main channels of dry mergers are between two elliptical galaxies or an ellitpical and a spiral galaxy. Dry mergers between spirals play almost no role and are a factor of 5 less frequent. The model predicts a large fraction of mixed mergers between a spiral and an elliptical galaxy. These are prime targets for the detection of dry mergers by tidal features \citep[e.g.][]{2005AJ....130.2647V,2008ApJ...684.1062F}.  In an earlier study \citet{2007MNRAS.381..389K}(K07) investigated the morphology of dry merger  progenitors in their SAM, finding that the majority is between two late-type galaxies in contrast to our results. There are two main reasons for this discrepancy between our models. While K07 use N-body simulations to follow the merging history of their model galaxies we apply the dynamical friction estimate to calculate the time it takes galaxies to merge once their haloes merge. It has been argued by e.g. K07, that this time scale is shorter than the actual merging time scale and hence would overproduce mergers. It is interesting to note, that the merger rate estimates based on the dynamical friction time scale in various SAMs or halo occupation models are in good agreement with the observations of the merger rate by \citet{jogee}. In contrast merger rates from SAMs based on N-body simulations following sub-haloes show systematic lower merger rates than the observations (Hopkins et al, in prep). Main problems in the sub-halo scheme for merging are too effective stripping of dark matter from the sub-haloes due to missing baryons and hence too long merging time scales, as well as calculating the merging time scale of the satellite galaxy once its hosting sub-halo has fallen below the halo resolution limit (Hopkins et al, in prep.). At this point it is still open which of the two schemes gives the more physical robust results. 
The second main reason for the differences between K07 and our results is the star formation efficiency. While we use a constant efficiency based on the local Schmidt-Kennicutt relation \citep{1998ApJ...498..541K} they use an efficiency that scales proportional to $M_{DM}^{0.73}$ for constant gas masses. As a consequence star formation for massive galaxies, which predominantly live in the most massive haloes, will be more efficient leaving them devoid of gas, but with massive stellar discs in their model. As seen in Fig. 10 of   \citet{2005ApJ...631...21K} their colour-magnitude relation shows an excess of very luminous blue galaxies, most likely associated with late-type galaxies, that subsequently take part in mergers. It should be noted however, that simulations of dry late-type mergers in general do not reproduce the kinematics and surface profiles of the most massive elliptical galaxies \citep[e.g.][]{2006MNRAS.369..625N} 
\begin{figure} 
\includegraphics[width=0.45\textwidth]{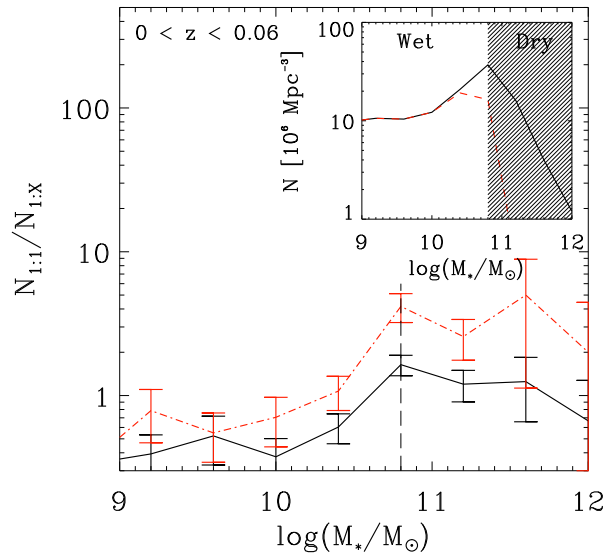}
 \caption{Fraction of 1:1 to 1:2 (solid line) and 1:1 to 1:3 (dot-dashed line) mergers as a function of {\bf remnant stellar mass within the redshift interval $ 0 < z < 0.06$. The error bars show Poisson error bars} Inset graph: Number density of all mergers (solid line) and of only wet mergers as a function of remnant mass within the same redshift interval. The black vertical dashed line indicates $M_{*,crit}$ }\label{fig5}
\end{figure}
\section{A New Way to Detect Dry Mergers}
In the last section, we argued that it is not straightforward to measure the dry merger rate from the evolution of the luminosity function. We here want to propose a novel approach that is observationally well accessible.  In Fig. 
\ref{fig5} we show the ratio of 1:1 mergers to 1:2 mergers and 1:3 mergers as a function of remnant galaxy mass. Here we count the number of mergers, dry and wet, with different mass ratios that occurred within the redshift interval $  0 < z < 0.06$.  Galaxies with masses below $M_{*} < M_{*,c}$ do not show any strong variation in their relative merger rates. Only on going to more massive galaxies does the relative contribution from different mass ratios start to  change. The variation in the merger rate is most pronounced at a mass scale of $M_{*,c}$, where equal mass mergers dominate the overall merger rate. Galaxies participating in these mergers generally live in halos with masses that are above $M_{DM,crit}$. The sudden increase in equal mass mergers is a direct consequence of the  shut-off mass scale $M_{DM,crit}$. Galaxies that just reached $M_{*,c}$ do not grow by star formation anymore: their main channel of growth is mergers. What happens is that galaxies grow until they reach $M_{*,c}$ and then stall until they merge with another galaxy of similar mass. Once galaxies passed $M_{*,c}$ the relative fractions of equal and unequal mass mergers approach the values below $M_{*,c}$. Another signature of the shut-off mass scale is imprinted in the overall number of dry and wet mergers  as a function of galaxy mass (see inset graph of Fig. \ref{fig5}). At $M_{*,c}$ the number density of dry mergers gets enhanced because of galaxies growing till they reach $M_{*,c}$ and then waiting to merge. The contribution of wet mergers towards the number density of all mergers drops very steeply at masses larger $M_{*,c}$ and allows us to clearly separate the dry merger activity region. We find a peak in the number density of mergers around $M_{*,c}$, which is around the same scale reported in a study of pair counts by 
\citet{2008arXiv0806.0018P}.
 
\section{Conclusion}
 In this Letter, we predicted properties of dry mergers in a model that assumes a critical shut-off mass scale for cooling of gas. The impact on the galaxy population and the merger rates can be summarized as follows. The high-mass end of the luminosity function is dominated by continued dry mergers. 
At any redshift $z \le 1$,  $10 \% - 20 \%$ of massive galaxies have had experienced a dry merger within their last Gyr. We find a dry merger rate of $ \sim 6 \times 10^{-5}$ Gyr$^{-1}$ Mpc$^{-3}$ and that the  number density of dry major mergers is significantly increased with respect to a model without shut-off mass scales. The relative fraction of equal mass mergers is enhanced with respect to unequal mass mergers at $M_{*,c}=6.3 \times 10^{10}$ M$_{\odot}$ which marks the transition of galaxies from being predominantly formed in gaseous mergers or through star formation in discs to dry mergers.  In a model where the transition between star forming and non-star forming galaxies is regulated by a 
physical process that does not result in a sharp shut-off mass scale, the relative rates of equal to unequal mass mergers do not show a significant change with mass (e.g. K07). 
Around the same mass scale, the merger rate is enhanced with respect to the general trend of a decreasing merger rate with mass consistent with recent observations by \citet{2008arXiv0806.0018P}. All of these features can be explained by considering that galaxies grow through star formation in discs only until their host halos reach $M_{DM,crit}$ and their supply of fuel in the form of cold gas stops. At this mass scale, efficient shock heating kicks in,
 as well as the gas becoming  prone to efficient heating from various feedback sources \citep{2006MNRAS.368....2D}. Galaxies on average have masses of $M_{*,c}$ when this is the case, and can only grow to become more massive by mergers. 
 As a result the relation between central galaxy stellar mass and host dark mater halo mass will become shallower, resulting in unequal mass dark halo mergers resulting in similar mass galaxy mergers (see also Hopkins et al. in prep).

The results presented here can be used to test the existence of a shut-off mass scale (see e.g. \citep{2009ApJ...695..900Y} for a recent study based on galaxy group catalogue, arguing that this mass scale must be larger than $10^{12.5}$ M$_{\odot}$). If indeed various physical processes conspire to generate a characteristic mass scale,  it should leave its fingerprint in the equal mass  merger rate. Using systematic surveys of galaxies to count pair statistics one can measure the relative fraction of equal to unequal mass mergers and look for a change as a function of mass. This approach is rather insensitive to difficulties with observing changes in luminosity functions, morphologies of galaxies or signs of interactions, and therefore should be able to provide robust results even at larger redshifts.
\\

We would like to thank the referee for his valuable comments, as well as Shardha Jogee, Gary Mamon, Michael Brown  and Avishai Dekel for helpful comments that improved the manuscript.



\label{lastpage}

\end{document}